\documentclass[11pt,dvips]{article}

\usepackage{epsfig,times}
\usepackage{picinpar}
\usepackage{wrapfig}
\usepackage{floatflt}
\makeatletter
\renewcommand{\section}{\@startsection{section}{1}{0in}
	{0.4\baselineskip}{0.1\baselineskip}{\Large\bf}}
\renewcommand{\subsection}{\@startsection{subsection}{2}{0in}
	{0.25\baselineskip}{-\baselineskip}{\large\bf}}
\renewcommand{\subsubsection}{\@startsection{subsubsection}{3}{0in}
	{0.1\baselineskip}{-\baselineskip}{\normalsize\bf}}
\makeatother
\begin{document}

%
\makeatletter\newcommand{\ps@icrc}{
\renewcommand{\@oddhead}{\slshape{OG 4.4.16}\hfil}}
\makeatother\thispagestyle{icrc}
%
\markright{OG 4.4.16}
\begin{center}
%
{\LARGE \bf The Washington Large Area Time Coincidence Array}
\end{center}

\begin{center}
%
%
{\bf E. Zager, J.G. Cramer, S.R. Elliott, J.F. Wilkerson,
 and R.J. Wilkes}\\
{\it Department of Physics, University of Washington, Seattle, WA 98195, USA\\
}
\end{center}

\begin{center}
{\large \bf Abstract\\}
\end{center}
\vspace{-0.5ex}
%
%
The number and density of schools in the Seattle area is convenient for the
study of distributed particle showers produced at the top of the atmosphere
by ultra-high energy ($>10^{19}$ eV) cosmic rays. We are forming a
collaboration
for the development of a distributed detector network
to study air showers from such ultra-high energy cosmic rays. We call the
cosmic ray
measurement component WALTA (WAshington Large-area Time-coincidence Array).
WALTA aims to provide teachers and students the opportunity to become active
participants in forefront scientific projects. A cornerstone of the program
will be to install a measurement module at each participating school.
%

\vspace{1ex}

%
%
\section{Introduction:}
\label{intro.sec}
	Ultra-high energy cosmic ray particles with energies greater than
about $5 \times 10^{19}$ eV cannot traverse great distances through the
cosmos without
losing energy to pion photoproduction in collisions with the cosmic
microwave background. Such a particle will lose energy until it is
below threshold for this reaction. This process, referred to as the
GZK cutoff (Greisen, 1966, and Zatsepin \& Kuzmin, 1966), happens in
about 50-100 Mpc (Cronin, 1992 and Aharonian \& Cronnin, 1994).
 Hence for a $10^{20}$ eV particle to arrive at Earth, it must
originate in the local neighborhood. The observed arrival direction of
  events with energies greater than $5 \times 10^{19}$ eV  provide no
compelling
evidence for a nearby source. However there
is a possible indication of correlations in event arrival direction (Cronin
1999).
Further, there are no known nearby astrophysical objects which
are likely to accelerate particles to these energies (Hillas,
1984 and Biermann, 1997). This paradox
is one of the major current mysteries in astrophysics.

\section{NNODE and WALTA:}
\label{NNODE.sec}
We are forming a collaboration for the development of a
distributed detector network that will support measurements of air
showers from ultra-high energy cosmic rays and can also support a broad class
of other physical measurements. Other researchers have expressed interest in
using the framework, for example,
 as a seismograph network for
geophysics studies, a passive radar network for ionosphere studies, and air
sampling
for atmospheric studies.  We call the overall network NNODE (Northwest Network
for Operation of Distributed Experiments), and we call the cosmic ray
measurement
component WALTA (WAshington Large-area Time-coincidence Array).  The focus
of the efforts of our subgroup within NNODE is WALTA.
The project is to be a direct physical science
outreach program between faculty and students of the University of Washington
and the science teachers and students of Washington-area schools (grades 7-12).
 The detection techniques and philosophy
 are similar to those of the ALTA project in Alberta, Canada
(Pinfold). The groups in Washington and Alberta
have a loose collaboration.

Technological advances of the past decade have made such distributed
experiments practical. Powerful yet inexpensive desktop computers are
available and each detection site can have its own unit. The Global
Positioning System (GPS) can provide very good absolute timing (Berns
99) and positioning between distributed sites. Finally the Internet
provides convenient communication between each of the sites.  Each
WALTA/NNODE measurement module is envisioned to consist of a computer
with an Internet connection, a GPS timing system, and measuring
equipment.  The measuring devices used will depend on the science
investigation.  For the WALTA investigation, modules will consist of
scintillation paddles to detect distributed particle showers produced
by ultra-high energy cosmic rays. We plan to ultimately locate the
majority of the modules at public and private schools and to enlarge
the network as additional schools participate.

An aspect of this program is that the teachers, and through them their
students, will have the opportunity to directly contribute in cutting
edge scientific research.  We expect that this direct involvement will
motivate and energize teachers and result in their learning science at
a deeper level.  It may also spark the imaginations of their students,
encouraging their consideration of careers in the sciences.  The
teachers and their students will be able to learn about scientific
techniques, mathematical tools, and the latest measurement technology
and see them applied as new results emerge.  Many elements of the
project will provide a basis for classroom teaching units in science,
mathematics, and technology.  The classroom learning will be made more
immediate by the ongoing measurements of the systems and the emergence
of new results.

\section{WALTA and Cosmic Ray Studies:}
\label{WALTA.sec}
An ultra-high energy cosmic ray colliding with the upper atmosphere will
produce
a cascade of particles which at ground level contains
several billion particles and covers tens of km$^2$.  Sampling
the particles from such a shower requires a distributed detector with sites
separated on the order of a km.

To study the potential performance of WALTA, we chose 32 candidate sites in the
Seattle area. (See Fig. 1) These locations are associated with the
University of Washington,
high schools, middle schools, and other local universities and community
colleges.
The list is neither exhaustive nor confirmed. At each site we anticipate
placing
three or four 1-inch thick scintillators of approximate area 1 m$^{2}$
each. These detectors
would be separated by approximately 10-20 m. At this spacing, the
individual sites will measure
cosmic rays of energy near the knee ($\approx 10^{15}$ eV). This will
provide students with high rate
data similar in character to the low rate data from ultra high energy events.

Each site would trigger independently.
We would require that some number of detectors
at each site observe a signal corresponding to a minimum number of vertical
equivalent muons. The data from
each site would be transferred via the Internet to a central computing
location
to be analyzed off-line
for inter-site coincidences.

The 32-site array covers a large area, approximately 200 km$^{2}$. However,
the spacing
is not uniform. The mean nearest-neighbor spacing between these sites is
approximately
1.4 km but varies greatly. As a result, ``holes" in the array reduce the
trigger
efficiency significantly. However, the autonomous nature of the individual
sites makes
the array very easy
to expand. That is, holes can be filled and the array can easily be
extended into
the urban areas north, east, and south of Seattle. Thus the array can be
made more
efficient and larger.

Many of the site and array parameters must be optimized. This optimization
is being studied
through simulation using the CORSIKA code, version 5.624 (Heck, 1998).
Figure 1 shows the site
layout with a simulated vertical 10$^{19}$ eV event to provide some
indication of the array response.
In the figure, the squares
are centered on sites which have no response to the event. The 5 sites
responding to the
event are indicated by circles. The area of
each circle is proportional to the number of particles detected at the
site. The number
of detectors
and their geometry, the detector spacing, and the trigger requirements are
all parameters
which are currently
under study. Furthermore, site locations required to eliminate holes in the
effective
detector sensitive area are being identified. Subsequently, we will search
for new candidate sites to fill such holes.

One very dangerous possibility is that
of high-energy tails on the resolution function. The observation of the
highest energy
cosmic rays is the critical feature of the experiment. If low energy events
below the
GZK cutoff can be falsely identified as having a high energy, it will
seriously compromise
the conclusions one can draw from the measurements. Simulations to study
this possibility
are also underway.

Finally efforts are underway to finalize the hardware design. To ensure the
ease
of expansion, the cost of each site must be minimized. Prototypes of a site
system are
being assembled and characterized.

\section{Conclusion:}
\label{conclusion.sec}

The scientific activity of the WALTA project has a number of aspects which can
feed into teacher and student research experience.  The technological examples
provided by the network of computer systems, the display of data as it is
acquired, the use of GPS circuits for timing, and the electronics of the
detection system will serve as challenging examples for the technologically
inclined students and the teachers of technology-based classes.
The analysis and interpretation of the results from many of the measurements
will provide case studies for the application of statistics and mathematics
that can be linked to mathematics classes.  The physics that underlies the
research activity also has many potential connections that can enrich high
school physics classes.

We wish to acknowledge the support and encouragement of Dean David Hodge,
College of Arts and Sciences, University of Washington. We also thank Steve
Ellis and David Boulware for their encouragement and advice. These efforts are
supported in part by the Department of Energy through the University of
Washington
Nuclear Physics laboratory.

%
\begin{figure}[bp]
\centering
\epsfig{figure=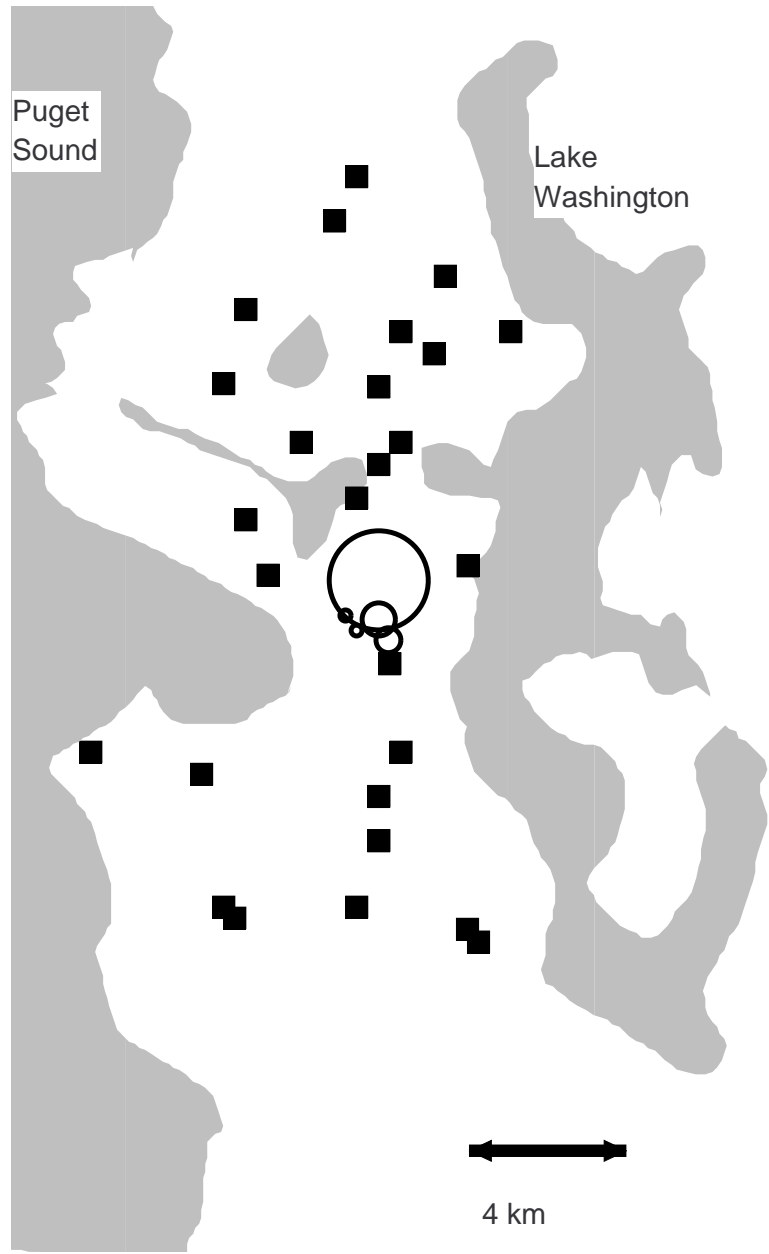,width=5in}
\vspace{0.7in}
\caption[WALTA Layout]
{The Seattle area indicating a number of suggested sites.}
\label{walta-layout}
\end{figure}
%
%
%
%
%
%
\vspace{1ex}
\begin{center}
{\Large\bf References}
\end{center}
%
Aharonian, F.A. \& Cronin, J.W., 1994, Phys. Rev. {\bf D50}, 1892.\\
Berns, H. G. \& Wilkes, R. J., 1999, Proc. 11th IEEE NPSS Conference.
Biermann, P., 1997, J. Phys. G {\bf 23}, 1.\\
Cronin, J.W., 1992, Nucl. Phys. B (Proc. Supp.) {\bf 28B}, 213.\\
Cronin, J.W., 1999, Rev. Mod. Phys. {\bf 71}, 165.\\
Greisen, K., 1966, Phys. Rev. Lett. 16, 748.\\
Heck, D., {\it et al.}, 1998, Karlsruhe Report FZKA 6019;
   http://www-ik3.fzk.de/~heck/corsika/Welcome.html \\
Hillas, A.M., 1984, Astron. Astrophys. {\bf 22}, 425.\\
Pinfold, J.L., {\it et al.}, see http://csr.phys.ualberta.ca/~alta for
proposal text.\\
Zatsepin, G.T., \& Kuzmin, V.A., 1966, JETP Lett. {\bf 4}, 78.\\

\end{document}